\documentstyle[12pt]{article}
\newcommand{\be}{\begin{equation}}
\newcommand{\ee}{\end{equation}}
\newcommand{\bear}{\begin{eqnarray}}
\newcommand{\ear}{\end{eqnarray}}

\date{}

\newcommand{\grgl}{\:\hbox to -0.2pt{\lower2.5pt\hbox{$\sim$}\hss}
           {\raise3pt\hbox{$>$}}\:}
\newcommand{\klgl}{\:\hbox to -0.2pt{\lower2.5pt\hbox{$\sim$}\hss}
           {\raise3pt\hbox{$<$}}\:}

\begin{document}
\begin{titlepage}
\begin{center}
{\bf\LARGE Time Evolution of Non-Equilibrium Effective Action}\\
\vspace{1cm}
Christof Wetterich\footnote{This work was performed
in part at ITP, UCSB, Santa Barbara, and supported in part by the 
National Science Foundation under Grant No. PHY94-07 197}\\
\bigskip
Institut  f\"ur Theoretische Physik\\
Universit\"at Heidelberg\\
Philosophenweg 16, D-69120 Heidelberg\\
\vspace{1cm}
\end{center}
\begin{abstract}
The time evolution of correlation functions in 
statistical systems is described by an exact
functional differential equation for the corresponding
generating functionals. This allows for a systematic 
discussion of non-equilibrium physics and the approach to 
equilibrium without the need of solving the nonlinear
microscopic equations of motion or computing the time
dependence of the probability distribution explicitly.\vspace*{0.8cm}
\end{abstract}
\end{titlepage}
A description of the time evolution of statistical systems is needed
in very diverse areas of science, ranging from the growth of density
fluctuations in the early universe to the evolution of a population
of bacteria. As a typical problem we may consider 
a given distribution of density fluctuations in the early universe
at the time when the photons
decouple - say a Gaussian distribution around
central values given by a flat Harrison-Zeldovich
spectrum - and ask how it looks like when the galaxies
form. For small fluctuations the underlying gravitational
equations of motion (or field equations) can be linearized
and approximate solutions found \cite{1}. For large
density contrasts, however, the nonlinearities in the field
equations become important and the analytical discussion
becomes more involved. Most approaches to this problem
therefore use numerical simulations. 
The observed galaxy correlation function shows
a power-like decay and similar for clusters.
This raises the challenge to understand the exponents
analytically - perhaps by the attraction of the 
nonlinear system to a suitable partial fixed
point, similar to the renormalization flow of
couplings in the theory of critical phenomena.

Examples from particle physics
include the rate of baryon number-violating
processes at high temperature and the discussion
of a possible formation of a disordered chiral
condensate \cite{2} in heavy ion collisions. After such a
collision the distribution of pion fields is not given by the
equilibrium distribution corresponding to the ground state. The 
time evolution of the non-equilibrium distribution could 
produce spectacular coherent effects. In statistical physics
one may ask how a system evolves to equilibrium after a
quench or starting in a metastable state. We propose here a 
method how such questions can be addressed
directly on the level of correlation functions without solving 
the (nonlinear) equations of motions or computing
the time evolution of the probability distribution. It is based on an
exact functional differential equation for the time evolution
of generating functionals for the correlations functions.
The microscopic laws or equations of motion reflect themselves
in the precise form of this evolution equation. Many general
aspects, as the approach to possible fixed points for $t\to\infty$,
are, however, independent of this precise form.

Consider a system of degrees of freedom $\chi_m(t)$ whose
time evolution is determined by a differential equation of
motion
\be\label{1}
\frac{d}{dt}\chi_m=\dot\chi_m=F_m[\chi]=
\sum^\infty_{k=0}f^{(k)}_{mn_1...n_k}\chi_{n_1}...\chi_{n_k}\ee
(We always sum over repeated indices and use $\chi_{-n}
=\chi^*_n$ with a summation over positive and negative values of
$n$ if $\chi_n$ is a complex variable. In case of continuous field 
equations for $\chi(x,t)$ the index summation corresponds to 
an integration over $x$ or a corresponding momentum summation
for Fourier modes. The equation of motion is nonlinear unless
$f^{(k)}$ vanishes for $k>1$ and $F_m$ can also
depend on $t$.) We assume that with given initial values
$\chi^0_m=\chi_m(t_0)$ eq. (1) has a solution $\chi_m(t;\chi^0)$, 
but we will not need its explicit form. The initial probability
distribution $\sim\exp-S_0[\chi^0]$ at time $t_0$ describes an ensemble
of initial conditions. The correlation functions for the variables
$\chi_m$ at some time $t>t_0$ are then given by
\be\label{2}
<\chi_{n_1}(t)...\chi_{n_i}(t)>=Z^{-1}\int D\chi^0\ \chi_{n_1}(t;\chi^0)...\chi_{n_i}(t;\chi^0)\exp-S_0[\chi^0]
\ee
with $Z=\int D\chi^0\exp-S_0[\chi^0]$ and $\int D\chi^0
\equiv\int\prod_md\chi^0_m$. Inserting the equation of motion
we see that the evolution of the mean value of $\chi_m$ involves
the higher correlation functions
\be\label{3}
\frac{d}{dt}<\chi_m(t)>=\sum^\infty_{k=0}f^{(k)}_{mn_1...n_k}<
\chi_{n_1}(t)...\chi_{n_k}(t)>\ee
Writing down similar equations for the correlation functions 
one arrives at an infinite system of differential equations for
the time dependence of the $n$-point functions.

Insight in the behaviour of this system can be gained
from the generating functional for time-dependent $n$-point
functions
\be\label{4}
Z[j,t]=\int D\chi^0\exp\{-S_0[\chi^0]+j^*_m\chi_m(t;\chi^0)\}\ee
or the related functionals for the connected correlation functions
$W[j,t]=\ln Z[j,t]$ or the 1PI irreducible correlation functions
\bear\label{5}
&&\Gamma[\varphi,t]=-\ln Z[j,t]+j^*_m\varphi_m,\nonumber\\
&&\varphi_m=\frac{\partial\ln Z[j]}{\partial j^*_m}\ear
We will concentrate here on the time evolution of the non-equilibrium
effective action $\Gamma[\varphi,t]$. Knowledge of $\Gamma[\varphi,
t]$ contains all information on the time evolution of correlation
functions. In particular, the mean value $<\chi_m(t)>$ corresponds
to the (time-dependent) value of $\varphi_m$ for which $\Gamma$
has its minimum. We should emphasize that in contrast to the equilibrium
effective action the non-equilibrium effective action
$\Gamma[\varphi,t]$ depends on the specific probability distribution
for the initial values $\chi^0$ at $t=t_0$, as
specified by $S_0[\chi^0]$ or,
equivalently, by $\Gamma[\varphi,t_0]$. Nevertheless, some features,
as the approach to a possible equilibrium state for $t\to\infty$, can
become independent of the initial conditions.

The time dependence of $\Gamma[\varphi,t]$ is specified by an exact
evolution equation ($\partial_t$ is the time derivative at fixed
$\varphi$)
\be\label{6}
\partial_t\Gamma[\varphi]=-\frac{\partial\Gamma[\varphi]}
{\partial\varphi_m}
\hat F_m[\varphi]\ee
\bear\label{7}
&&\hat F_m[\varphi]=f_m^{(0)}+f_{mn}^{(1)}\varphi_n+f^{(2)}_
{mn_1,-n_2}\{\varphi_{n_1}\varphi^*_{n_2}+(\Gamma^{(2)})^{-1}_{n_1n_2}\}
\nonumber\\
&&+f_{mn_1n_2,-n_3}^{(3)}\{\varphi_{n_1}\varphi_{n_2}\varphi^*_{n_3}
+\varphi_{n_1}(\Gamma^{(2)})^{-1}_{n_2n_3}+
\varphi_{n_2}(\Gamma^{(2)})^{-1}_{n_1n_3}+\varphi_{n_3^*}(\Gamma^{(2)})
^{-1}_{n_1,-n_2}\nonumber\\
&&-
(\Gamma^{(2)})^{-1}_{n_1p_1}(\Gamma^{(2)})^{-1}_{n_2p_2}
(\Gamma^{(2)})^{-1}_{p_3n_3}\frac{\partial^3\Gamma}
{\partial\varphi^*_{p_1}\partial\varphi^*_{p_2}\partial\varphi_{p_3}}
\}\nonumber\\
&&+\sum^\infty_{k=4}f^{(k)}_{mn_1...n_{k-1},-n_k}
(\varphi_{n_1}+(\Gamma^{(2)})^{-1}_{n_1p_1}\frac{\partial}{\partial
\varphi^*_{p_1}})\nonumber\\
&&...(\varphi_{n_{k-2}}+(\Gamma^{(2)})^{-1}_{n_{k-2}p_{k-2}}
\frac{\partial}{\partial\varphi^*_{p_{k-2}}})(\varphi_{n_{k-1}}
\varphi^*_{n_k}+(\Gamma^{(2)})^{-1}
_{n_{k-1}n_k})\ear
Here the matrix $\Gamma^{(2)}[\varphi]$ is given by the second
derivatives
\be\label{8}
(\Gamma^{(2)})_{mn}=\frac{\partial^2\Gamma[\varphi]}{\partial
\varphi^*_m\partial\varphi_n}\ee
and its inverse $(\Gamma^{(2)})^{-1}$ denotes the effective
($\varphi$-dependent) propagator. Eq. (6) is the central 
equation of this letter and we observe that it
simplifies considerably if the equation of motion contains
only up to quadratic $(k=2)$ or cubic $(k=3)$ terms. In the 
limit of infinitely many degrees of freedom $\chi_m(t)\equiv
\chi(x,t)$, it is a functional differential equation and $\partial
/\partial\varphi_m$ may be replaced by $\delta/\delta\varphi(x)$.
Eq. (6) follows from the evolution equation for $Z[j,t]$ (with
$\partial_t$ at fixed $j$ here)
\be\label{9}
\partial_tZ[j]=j^*_mF_m[\frac{\partial}{\partial j^*}]
Z[j]\ee
and the identity $\partial_t\Gamma_{|\varphi}=-\partial_t\ln Z_{|j}$.
In turn, eq. (9) can be proven by taking a time derivative of eq. (4), 
inserting the equation of motion (1) under the (functional) integral and 
expressing the powers of $\chi_{n_i}(t)$ by derivatives with
respect to $j^*_{n_i}$.

The evolution equation for the mean value $\bar
\varphi_m(t)=<\chi_m(t)>$ obtains from the minimum condition
$(\partial\Gamma/\partial\varphi_m)(\bar\varphi(t))=0$ which is
valid for all $t$ 
\be\label{10}
\frac{d}{dt}\bar\varphi_m(t)=\hat F_m(\bar\varphi(t))\ee
Comparing the ``macroscopic'' equation of motion (10) 
which includes the fluctuations with the microscopic one
(1) for a particular configuration, we see that the influence
of the fluctuations appears through the terms in $\hat F_m$
(7) which involve the propagator $(\Gamma^{(2)})^{-1}$ and 
those involving higher $\varphi$-derivatives of $\Gamma$.
The mean values of the connected two-point correlation
function $<\chi_m(t)\chi_n^*(t)>-<\chi_m(t)>
<\chi^*_n(t)>\\
=(\Gamma^{(2)})^{-1}_{mn}(\bar\varphi(t))$ evolves according
to
\bear\label{11}
&&\frac{d}{dt}P_{mn}\equiv\frac{d}{dt}\Gamma^{(2)}_{mn}
(\bar\varphi(t))=A^*_{pm}P_{pn}+P_{mp}A_{pn}\nonumber\\
&&A_{mn}=-\frac{\partial\hat F_m}{\partial\varphi_n}(
\bar\varphi(t))\ear
In the limit where $\partial\hat F_m[\varphi]/\partial\varphi_n$ 
is independent of $\varphi$ (in particular for a linear 
equation of motion where $A_{mn}=-f^{(1)}_{mn}$) one can solve
eq. (6) explicitly. The time evolution of all 1PI $n$-point 
functions is given by appropriate contractions with $A$ similar
to eq. (11) and therefore determined by the eigenvalues of $A$. 
For small nonlinearities one can perturbatively expand around 
this solution. Of course, the magnitude of the nonlinear effects
is not only determined by the values of the coefficients
$f^{(k)}$ but also by the size of the fluctuations.

Let us next come to the interesting question if the evolution
equation (6) has fixed points $\Gamma_*[\varphi]$ which are
approached for $t\to\infty$. Such fixed points could correspond
to possible equilibrium distributions to which the system
is attracted. The existence of fixed points (solutions of 
$\partial_t\Gamma[\varphi]=0$) obviously depends on the precise
form of the equation of motion (1). We concentrate in the following
on conservative equations of motion
\be\label{12}
\dot\chi_{1s}=B_{st}\frac{\partial U}{\partial\chi^*_{2t}},\quad
\dot\chi_{2s}=-B_{st}^\dagger\frac{\partial U}{\partial\chi^*_{1t}}
\ee
which can be derived from some generalized potential $U
[\chi]$ which is conserved if $U$ does not 
depend explicitly on time $(dU/dt=\partial_tU_{|\chi})$. 
For conserved $U$ one can show that
\be\label{13}
Z_*[j,\beta]=\int D\chi\exp(-\beta U[\chi]
+j^*_m\chi_m)\ee
is a fixed point of eq. (9). (Here $m=(1,s),(2,s)$ is a
collective index.) The Legendre transform (5) $\Gamma_*[\varphi,
\beta]$ is then a fixed point for the evolution equation (6) for
arbitrary values of $\beta$. The proof relies on the invariance
of the measure $D\chi$ under variable shifts
$\chi_m'=\chi_m+\epsilon_m$
wich constant (infinitestimal) $\epsilon_m$. Expanding
\be\label{14}
U[\chi]=\sum^\infty_{k=0}\frac{1}{k!}u^{(k)}_{n_1...n_k}
\chi_{n_1}\chi_{n_2}...\chi_{n_k}\ee
the shift invariance implies the Schwinger-Dyson equation \cite{3}
\be\label{15}
\beta\sum^\infty_{k=0}\frac{1}{k!}u^{(k+1)}_{n_1...n_k,-m}
Z^{-1}_*[j]\frac{\partial}{\partial j_{n_1}^*}...
\frac{\partial}{\partial j_{n_k}^*}Z_*[j]=j_m\ee
The coefficients $f^{(k)}$ in the equation of
motion (1) are directly related
to $u^{(k+1)}$ and the right-hand side of eq. (9) vanishes
due to the relative minus sign in (12).

The symplectic structure of eq. (12) is typical for second-order
equations of motions
\be\label{16}
\ddot\chi_m=-\frac{\partial V[\chi]}{\partial\chi^*_m}\ee
Identifying $\chi_1$ with $\chi$ and $\chi_2$ with $\pi
\equiv\dot\chi$ the equation of motion takes the form (12) 
with $B_{st}=\delta_{st}$ and
\be\label{17}
U=V[\chi]+\frac{1}{2}\pi_m^*\pi_m\ee
We interpret $U$  as the total energy of the system and $\beta$
as the inverse temperature. The ``kinetic part''
of $Z_*$ which involves the sources for $\pi_m$
factorizes as a trivial
Gaussian. The remaining part of $Z_*$ is the partition function
for a system $\chi_m$ with (potential) energy $V[\chi]$ in
thermodynamic equilibrium at temperature $1/\beta$. We note
that the concept of temperature arises here without any
notion of interaction with some external heat bath! Since $U$
is conserved and $<U>=-\partial\ln Z_*[j=0]/\partial\beta$
the temperature is uniquely fixed by the initial value
$<U>[t_0]$ if $Z_*[0,\beta]$ is monotonic
in the appropriate range.

If the thermodynamic equilibrium partition function $Z_*$ were 
the only fixed point of the system and attractive for
$t\to\infty$, we could  be sure that initial conditions not too far
from equilibrium would ``thermalize'', i.e. $\Gamma[\varphi,
t\to\infty]=\Gamma_*[\varphi]$. The approach to the equilibrium
fixed point may, however, be obstructed by the existence of conserved 
quantities $A_i$ besides $U$. The fixed point can be reached
only if the initial values of all $A_i$ coincide
with the equilibrium values. The generic situation may be similar
to quantum mechanics where the trace of arbitrary powers
of the density matrix is conserved. In what sense
(reduction of effective degrees of freedom, time averaging)
an equilibrium state can be reached under such circumstances
is an interesting and subtle question.

Let us finally turn to more practical applications
of eq. (6) to  non-equilibrium situations. Exact solutions are 
out of reach and one has to ressort to approximations
by truncating the most general form of $\Gamma[\varphi,t]$
to an object of manageable size. The truncation has to be adapted
to the particular system and the success of the method in the
nonlinear non-equilibrium domain will depend on the ability
to conceive a realistic truncation. As an example we consider
the $O(N)$ symmetric $\varphi^4$-theory
in $d$ dimensions, with equation of motion $(i=1..d, a=1...
N,\chi_a(x,t)$ real)
\be\label{18}
\ddot\chi_a(x,t)=\partial_i\partial_i\chi_a(x,t)
-m^2\chi_a(x,t)-\frac{1}{2}\lambda
\chi_b(x,t)\chi_b(x,t)\chi_a(x,t)\ee
For $\lambda\to\infty$ with fixed negative $m^2/\lambda$ this
corresponds to Ising or Heisenberg models, and for 
$d=3, N=4$ and $m^2<0$ this is the equation used in
numerical simulations of pion dynamics related to the 
investigation of a possible disordered chiral condensate. For the non-equilibrium effective action we use the
truncation
\bear\label{19}
\Gamma[\varphi,t]&=&\frac{1}{2}\beta\int\frac{d^dq}{(2\pi)^d}
\{A(q)\varphi^*_a(q)\varphi_a(q)+B(q)\pi^*_a(q)
\pi_a(q)+2C(q)\pi^*_a(q)\varphi_a(q)\}
\nonumber\\
&&+\frac{1}{2}\beta\int d^dx\{u\ \rho^2(x)+v\rho(x)\pi_a(x)\varphi_a
(x)\}\ear
with $\rho=\frac{1}{2}\varphi_a\varphi_a,\varphi_a(q)
=\int d^dxe^{iq_ix_i}\varphi_a(x)=\varphi_a^*(-q)$.
The evolution equation (6) describes the time dependence
of the real functions $A(q),B(q),C(q)$ which depend here 
only on the invariant $q_iq_i$, and the constants $u$ and $v$.
Inserting (19) in (6) and omitting on the r.h.s. terms $\sim v$,
one finds after some algebra
\bear\label{20}
\partial_tA(q)&=&2\omega^2_qC(q),\ \partial_tB(q)=-2C(q)\nonumber\\
\partial_tC(q)&=&\omega_q^2B(q)-A(q)\nonumber\\
\partial_tu&=&4\lambda C(0)K\nonumber\\
\partial_tv&=&-2u+2\lambda B(0)K\ear
with
\bear\label{21}
\omega_q^2&=&q^2+m^2+\frac{N+2}{2}\frac{\lambda}{\beta}
\Bigl(\int\frac{d^dq'}{(2\pi)^d}G(q')\nonumber\\
&&-\frac{u}{\beta}\int\frac{d^dq'}{(2\pi)^d}\frac{d^dq''}{(2\pi)^d}
G(q')G(q'')G(q+q'-q'')\Bigr)\nonumber\\
K&=&1-\frac{N+8}{2}\frac{u}{\beta}\int\frac{d^dq}{(2\pi)^d}
G^2(q)\nonumber\\
&&+\frac{3N+20}{2}\left(\frac{u}{\beta}\right)^2
\int\frac{d^dq}{(2\pi)^d}\frac{d^dq'}{(2\pi)^d}G^2(q)G(q')
G(q-q')\nonumber\\
G(q)&=&B(q)/(A(q)B(q)-C^2(q))\ear
We assume here that the system is regularized by a momentum
cutoff for $q^2$. For a given initial distribution specified
by $A,B,C,u$ and $v$ for $t=t_0$ the system (20) 
(or a simplified version with $B(q)=B(0), C(q)=C(0),
A(q)=A(0)+Zq^2$) could be
solved numerically. (Note that eq. (20) implies that $A(q)B(q)-C^2(q)$
is a conserved quantity for every $q$.)
Results may be compared with direct
numerical simulations which solve eq. (18) numerically and
average over initial conditions. In particular, 
the expectation value of a space-independent field obeys
$(\bar\rho=\frac{1}{2}\bar\varphi_a\bar\varphi_a)$
\be\label{22}
A(0)+u\bar\rho-\frac{1}{B(0)}(C^2(0)
+2vC(0)\bar\rho+\frac{3}{4}v^2\bar\rho^2)=0\ee
or $\bar\varphi_a=0$. It is easy to verify that the system
(20) has the equilibrium fixed point (13) with 
$C_*(q)=0,v_*=0,B_*(q)=1$ and $A_*(q)=\omega^2_q,
u_*=\lambda K$. The two last conditions are the (classical!)
Schwinger-Dyson equations for the propagator and the
quartic scalar coupling in the appropriate truncation. In thermodynamic
equilibrium the system exhibits spontaneous symmetry breaking
for $A_*(0)<0$. The contributions involving $C$ an $v$ on the
r.h.s. of eq. (22) describe the non-equilibrium deviation of
$\bar\rho$ from its equilibrium value $\bar\rho_*=-A_*(0)/u_*$.
Not too far from equilibrium $B(0)$ is positive and the
non-equilibrium effective mass term $A(0)-C^2(0)/B(0)$
is smaller than in equilibrium, thus enhancing the tendency
to spontaneous symmetry breaking.

We finally remark that the present formalism can be extended
to stochastic equations of motion and the description of correlation
functions at unequal time. This will permit to establish the
contact to path integral formulations \cite{4} 
and quantum field theory.


\begin{thebibliography}{10}
\bibitem{1}  E. Kolb and M. Turner, The Early Universe, 
Addison-Wesley (1990)
\bibitem{2} K. Rajagopal and F. Wilczek, Nucl. Phys.
{\bf B379} (1993) 395; {\bf B404} (1993) 577;\\
J. D. Bjorken, K. L. Kowalski and C. C. Taylor,
SLAC-PUB-6109 (1993)
\bibitem{3} F. J. Dyson, Phys. Rev. {\bf 75}
(1949) 1736;\\
J. Schwinger, Proc. Nat. Acad. Sc. {\bf 37} (1951) 452, 455
\bibitem{4} J. Zinn-Justin, Quantum Field Theory and
Critical Phenomena, Oxford University Press (1989)
\end{thebibliography}
\end{document}